\documentclass[12pt]{article}

\newcommand{\be}{\begin{equation}}
\newcommand{\ee}{\end{equation}}
\newcommand{\ba}{\begin{eqnarray}}
\newcommand{\ea}{\end{eqnarray}}

\begin{document}
\begin{center}
{\bf{On the pre-nucleosynthesis cosmological period}}
\vskip 0.5cm

R. Aldrovandi 
\vskip 0.2cm
Instituto de F\'{\i}sica Te\'orica \\ State University of S\~ao Paulo - UNESP \\ 
Rua Pamplona 145 \\ 01405 - 900 \ S\~ao Paulo \ Brazil
\vskip 0.5cm
J. Gariel and G. Marcilhacy
\vskip 0.2cm
Laboratoire de Gravitation et Cosmologie 
Relativistes \\ Universit{\'e} Pierre et Marie Curie, CNRS/ESA 7065 \\ 
4, Place Jussieu \ Cedex 05 \\ 75252 Paris, France
\end{center}
\vskip 0.5cm

\begin{abstract}
 Physics, as known from our local, around--earth experience, meets
 some of its applicability limits at the time just preceding the
 period of primeval nucleosynthesis.  Attention is focussed here on
 the effects of the nucleon size.  Radiation--belonging nucleons 
are found to produce an extremely high pressure at $kT \approx$ 
some tens or hundreds of $MeV$.  Quark deconfinement at higher 
energies would not change the results. 
\end{abstract}

%\pacs{98.80.Bp  ;  98.80.Cq.  }
 
%\maketitle

\section{Introduction}

The standard procedure of Physical Cosmology is to take present-day
knowledge and data and travel backwards in time, applying as well as possible our
local, laboratory- and observatory-tested Physics.  That Physics, as
we know it today, is able to explain so many of the progressively
distant and red--shifted data is the best mark we have of its
astounding range of validity.  The remotest time for which we have
reliable results is the nucleosynthesis epoch: well--established
Physics is able to give a fair account of the cosmological origin of
the lightest elements.  We shall here be interested in the period just
preceding that nucleosynthesis era, which we shall call
``pre--nucleosynthesis period'' (PNS period).

The end of that period -- the beginning of the nucleosynthesis age --
must correspond to a temperature $kT$ of the order of the deuteron
binding-energy, which means a few MeVs and a red-shift $z \approx$ 2
$\times 10^{10}$.  The PNS period could also be called the
``close-packing period''.  The ``closely-packed'' constituents
referred to are protons or, more precisely, nucleons.  A rough estimate
gives for their concentration a value around $10^{23} cm^{-3}$ and for
their mean free path $\lambda \approx $ $(n_{b}\sigma )^{-1}\approx
(10^{23} \times $ $10^{-26})^{-1}= 1000 \; cm$.  The usual
approximation assuming ideal fluids fails, but the 
physical assumptions on the large ratio between the total volume of
the system and the total volume occupied by the constituents are valid
and current Physics can be expected to hold.

Protons are remarkably stable (lifetime larger than $1.6 \times
10^{25}$ years~\cite{PDG98}), and neutrons decay into protons.  Thus,
we can safely suppose that the nucleons present today in the Universe
have been around from the ``beginning''.  The values of the critical
density and baryon density [see below, equation (\ref{range})] imply a
remnant nucleon density $n_{N}$ in the range $ 0.059 \leq n_{N} \leq $
0.296 $ (nucleon \times m^{-3})$ \ at present time.  Now, each nucleon
occupies a volume of the order $2.2 \times 10^{-45} m^{3}$, which
means that at $z \approx 10^{15}$ they attain a tightly packed state:
one nucleon per nucleon volume.  This will define for us the beginning
of the PNS period.  The mean free path is then of the order of the
size of the constituents.  In a nutshell: the causally-related
Universe has a volume $V_{U}$ $\approx$ $10^{81}cm^{3}$ and contains
$N_{N} \approx10^{74}$ nucleons.  The total ``internal'' volume of
these nucleons is $V_{n}\approx 10^{35} cm^{3}$.  The Universe had
that volume when $z \approx 10^{15}$.  The assumption of infinite
system volume -- which underlies the thermodynamic limit, as well as
the very definition of cross-section -- is then at least doubtful, and
the ideal fluid hypothesis is clearly untenable.  We shall later
refine this crude estimate, but the result will be, not quite
surprisingly, essentially the same for the remnant protons.  The
interest of the more refined approach rests on its formulas, which can
be applied to the protons belonging to the radiation bath.

The PNS period runs consequently between $z \approx 10^{10}$ and $z
\approx 10^{15}$.  The strategy to be followed will be rather
circular.  Quark deconfinement will be ignored to start with and
protons will be taken as stable.  We shall describe them by a potential and
find that the pressure related to present--day matter tends to an
infinite value.  That would happen, however, at energies for which the
notion of potential does not apply and for which deconfinement is
quite possible.  We then reconsider the question from the point of
view of the radiation--belonging nucleons, and find the same effect at
much lower energies, for which potentials do have a meaning and there
is no possibility of deconfinement.

In a first dealing with such unusual conditions we shall feel
justified in taking a na\"{\i}ve approach.  Instead of facing the
intricacies of the high-density matter equation of state \cite{Can},
we shall content ourselves with reasonable order--of--magnitude
estimates.  The proton will be considered as a hard constituent,
occupying an irreducible hard-core volume $\simeq $ $ 1 fermi^{3}$
$\simeq 10^{-39}cm^{3}$, represented by a hard-sphere potential. 
Potentials have been used from time to time in Cosmology, for instance
to show how the initial singularity can be avoided \cite{PW90}.  They
should, of course, be carefully handled in relativistic conditions. 
We shall be attentive to the energy conditions under which the very
notion of potential can loose its validity.

Section \ref{Friedmannmodel} is a summary of the Standard Model,
actually a commented formulary devoted to fixing notation, showing the
numbers we use and summing up some observation values relevant to our
subject.  We shall ignore non-standard possibilities, as eventual
``dark'' constituents, and accept usual reasonable assumptions, such
as the Debye screening which renders electrostatic effects negligible. 
Such a review of well--known topics is necessary to show how and when
we part from the standard procedure.  In section \ref{physics} a
general overview of physical problems appearing in the PNS period is
given.  In particular, we present our assumption that protons keep
their identity in energies much higher than usually supposed.  We then
proceed to a discussion of the hard-sphere potential and to a na\"{\i}ve
application to the Friedmann equations.  The result is that an
infinite matter pressure would block the backward progress at around
 $z \approx $ $10^{15}$ if we take into account only the ``remnant''
protons existing at present time.  At those red--shifts the nucleons
are relativistic and the idea of a hard-sphere potential is unrealistic, but it is
easier to argue starting from the consideration of the remnant
present--day protons.  We show then (section \ref{radionucleons}) that
the protons appearing through pair-creation from the radiation
background produce the same effect at much lower energies.  Pairs
of photons with energy barely enough to produce proton--antiproton
pairs will create non-relativistic protons and anti--protons, and for
these the notion of a potential does make sense.  Pair creation is a
very efficient process: the number of created protons is very large
already at  energies much lower than $1\ GeV$. The pressure blockage, in consequence,
 takes place at
rather low red--shifts.  The possible meanings of these results are
discussed in the last section.  A briefing on 
relativistic quantum gases is given in Appendix A.

%%%%%%%%%%%%%%%%%%%%%%%%%%%%%%%%%%
%					%
   \section{The standard model}	%
%					%
%%%%%%%%%%%%%%%%%%%%%%%%%%%%%%%%%%
\label{Friedmannmodel}

The large scale evolution of the Universe is described \cite{Wei72,CE97} by 
the two Friedmann equations for the scale parameter $a(t)$:

\begin{equation} {\dot a}^{2} = \left[ 2 \left( 
\frac{4 \pi G}{3} \right) \rho+ \frac{\Lambda c^{2}}{3} \right] a^{2} 
- k c^{2} \label{Friedmann1} \end{equation} % 
\begin{equation} {\ddot a} = \left[ \frac{\Lambda c^{2}}{3} - 
\frac{4 \pi G}{3} \left( \rho+ \frac{3 p}{c^{2}} \right) \right] a(t) 
 \label{Friedmann2} \end{equation} % 
which, once combined, lead to the two equivalent expressions 
\begin{equation} \frac{d \rho}{dt} = - 3 \ \frac{{\dot a}}{a} \ \left(\rho+
\frac{p}{c^{2}}\right) \; ,  \label{rhopoint} \end{equation} % 
  \begin{equation} %
\frac{d }{da} (\epsilon a^{3}) + 3 \ p \ a^{2} = 0 . 
\label{Friedmann4}
 \end{equation}
This equation  can be
alternatively obtained from the vanishing of the covariant divergence of 
the source energy-momentum tensor, and reflects simply
 energy conservation. 
Notation is hopefully obvious:  $\rho$ = $\epsilon/c^{2}$ is the source 
energy density in mass,
$p$ the pressure, $\Lambda$ the cosmological constant.  The
index
``0'' will indicate present--day values: the red-shift $z$, 
for example, is given by %
\begin{equation} %
1+z=\frac{a(t_{0})}{a(t)} \; . \label{zdefined}
 \end{equation} % %
The Hubble function %
\begin{equation} %
H(t) = 
\frac{{\dot a}(t)}{a(t)} = \frac{d}{dt} \ln a(t)  %
\end{equation} %%
has the present-day value $H_{0}=100\;h\; km\;sec^{-1}\;Mpc^{-1} $ =
$3.24 \times 10^{-18}\;h \; sec^{-1}$ (with the parameter $h$, of the
order of unity, encapsulating the uncertainty in present-day
measurements).  

Given an equation of state in
the form
$p$ =
$p(\rho)$, equation (\ref{Friedmann4}) can be integrated to give
\begin{equation} %
1 + z =\frac{a_{0}}{a(t)}= e^{\frac{1}{3} \int_{\epsilon_{0}}^{\epsilon}
\frac{d \epsilon}{\epsilon + p(\epsilon)}} \; . \label{onepluszee}
\end{equation} %
 For example, for a pure radiation content the equation of
state is $p$ = $\frac{1}{3} \epsilon$, so that
$$%
\epsilon_{z} = \epsilon_{0}\ (1+z)^{4} \ .
$$%
The energy density of dust matter, with $p = 0$, will behave according to 
$$%
\epsilon_{z} = \epsilon_{0}\ (1+z)^{3} \ .
$$%
Recall that Eq.(\ref{Friedmann4}) is a mere consequence of energy conservation.
These results are independent of the parameters $k$ and $\Lambda$. The relationship
between $z$ and the Hubble function is easily  found.
Taking the time derivative of Eq.(\ref{zdefined}) and comparing with the last
expression, one arrives at $\frac{dz}{dt}$ = $-\ H(t) (1+z)$, which integrates to
\begin{equation}
1 + z = e^{-\ \int_{t_0}^t H(t) dt} \ .\label{zandH}
\end{equation}

The critical mass density is
\begin{equation}%
\rho _{crit}=\frac{3H_{0}^{2}}{8\pi G}= 1.88 \times 
10^{-26}\;h^{2}\; \label{rhocrit}
kg \times m^{-3} \; .
\end{equation} %
 The baryon concentration and mass density are given by
\begin{equation}%
n_{b}=11.2\ \Omega_{b0}\;h^{2}\;(1+z)^{3}\;\; m^{-3}  \label{nb}
\end{equation} %
\begin{equation}%
\rho _{b}=1.88
\times 10^{-26} \Omega _{b0}\;h^{2}\;(1+z)^{3} kg\;m^{-3} ,
\label{rhob}
\end{equation} %
where the parameter \ $\Omega _{b0}$ = $\frac{8\pi G}{3}\frac{\rho _{b 
0}}{H_{0}^{2}}$ = $\frac{\rho _{b 0}}{ \rho _{crit}}$ \ has 
observational values in the range
\begin{equation}
0.0052 \ \leq \ \Omega _{b0}h^{2} \ \leq 0.026 \ . \label{range} %
\end{equation} %
These values for the critical  and baryon density lead to the 
remnant nucleon density  range used in the Introduction. 
In terms of $H(t)$, the equations can be written as %
\begin{equation} %
H^{2} = 2 \left( \frac{4 \pi G}{3}\right) \rho- \frac{k 
c^{2}}{a^{2}} + \frac{\Lambda c^{2}}{3 } \label{Htwo} %
\end{equation} % %
\begin{equation} %
{\dot H} = - \ 4 \pi\ 
G \left( \rho+ \frac{p}{c^{2}} \right) + \frac{k c^{2}%
}{a^{2}} \label{dotH} %
\end{equation} %
\begin{equation} %
{\dot a}(t) = H(t) \; a(t) \ ; \ %
\frac{d \rho}{dt} = - 3 H (\rho+ \frac{p}{c^{2}}) \; . 
\end{equation} % %
Introducing $\Omega _{m}$ = $\frac{\rho }{\rho _{crit}}$, 
$\Omega_{\Lambda 
}=\frac{\Lambda c^{2}}{3H_{0}^{2}} $ and $ \Omega_{k}(t)$ = $ 
-\;\frac{kc^{2}}{a^{2}H_{0}^{2}}$,  equation (\ref{Htwo}) takes the 
form %
\begin{equation} %
\frac{H^{2}}{H_{0}^{2}}=\frac{\rho }{\rho _{crit}}-\frac{kc^{2}}{
a^{2}H_{0}^{2}}+\frac{\Lambda c^{2}}{3H_{0}^{2}} = \Omega _{m}+\Omega
_{k}+\Omega _{\Lambda } \; . %
\end{equation} % 
Notice that $\Omega _{m}$ refers to the total 
amount of content: if baryons and radiation are to be considered, then 
$\Omega _{m}$ = $\Omega _{b}$ + $\Omega_{\gamma}$. The above expression
gives on present-day values the constraint  %
\begin{equation} %
\Omega_{m 0} + \Omega_{k0} + 
\Omega_{\Lambda} = \Omega_{b 0} + \Omega_{\gamma 0} + \Omega_{k0} + 
\Omega_{\Lambda} = 1.  \label{omegas} %
\end{equation} % %
We have used $\Omega_{k0}$ = $ -\;\frac{kc^{2}}{a_{0}^{2}H_{0}^{2}}$,
with $a_{0}$ = $a(t_{0})$.   There is a recent evidence for a large
value of $\Omega_{\Lambda}$ and a small value of
$\Omega_{k0}$~\cite{largelambda}.  Choosing for time and
length the convenient units %
\begin{equation} %
H_{0}^{-1}=3.0857 \times 10^{17}h^{-1} \; sec \; \  ; \ 
\frac{c}{H_{0}}=9.25 \times 10^{25}h^{-1} \; \; m \; , 
\label{cosmounits} 
\end{equation} % 
the  Friedmann  equations acquire the simpler forms
\begin{equation}
H^{2}=\frac{\rho}{\rho _{crit}}-\frac{k}{a(t)^{2}}+\frac{\Lambda}{3}
\label{H2}
\end{equation}
\begin{equation}
{\dot{H}}=-\ \frac{3}{2}\ H^{2}-\ \frac{3}{2}\ \frac{p}{c^{2}\rho 
_{crit}}\ +\ \frac{\Lambda }{2}-\ \frac{1}{2}\ \frac{k}{a(t)^{2}} \; .  
\label{Hpoint}
\end{equation} %

Gases at high energies in the presence of pair production are also better
considered in adapted units, which we introduce here while leaving
details to Appendix A. First of all, given a particle of mass $m$, it
is convenient to
use %
\begin{equation} %
\tau = \frac{kT}{mc^2}   \label{tau} \end{equation} % %
as the temperature variable. There are also two lengths of major interest, the 
Compton length and the thermal wavelength.  The static Compton length 
is a most natural unit of length: %
\begin{equation}%
\lambda_{C} = \frac{{\hbar} c}{m c^2} \; . 
\end{equation} % %
For the electron and for the proton, respectively, $\lambda_e$ $= 
3.81 \times 10^{-11}\ cm$  and $\lambda_p$ $= 2.08 \times 
10^{-14}\ cm$.  A natural volume cell for the proton will be 
$\lambda_p^3$ $= 9.0 \times 10^{-42}\ cm^3$.

If $\beta$ = $1/kT$ is the inverse temperature, (the cube of) the  relativistic
generaliza\-tion~\cite{AML83} of the thermal wavelength is given by %
\begin{equation} %
\Lambda_{T}^3(\beta) = 2 \ \pi^2  \beta  m c^2 \; \frac{e^{- \beta m 
c^2}}{K_2(\beta m c^2)} \; \left(\frac{\hbar c}{m c^2} \right)^3 = 
\frac{2 \ \pi^2}{ \tau} \; \frac{e^{- 1/\tau}}{K_2(1/\tau)} \; 
\lambda_{C}^3 \; .  \label{thermlength} \end{equation} % %
Here $K_{2}(x)$ is the modified Bessel function of second order, 
whose asymptotic behaviors, leading to the nonrelativistic and the 
ultra--relativistic limits,  are given in Appendix A. Here we only 
remark that the non-relativistic limit gives the usual expression %
\begin{equation} %
\Lambda_{NR}(\beta) = \lambda = \hbar \ \sqrt{\frac{2 \pi \beta}{m}} 
= \ \sqrt{\frac{2 \pi}{\tau}} \; \lambda_C\ .  \label{lambdaNR} %
\end{equation} % %
For instance, a proton at $k T \approx$ $4 \ MeV$ will have 
$\Lambda_{NR}$ $\approx$ $40 \ \lambda_p$.  A  proton
will ``occupy'' a degeneracy-volume $\lambda^{3}$ = $\frac{1.42 \times 
10^{-40}}{\tau^{3/2}} \; cm^3$, from which every other proton will be 
statistically excluded by Fermi repulsion.  The ultra-relativistic limit will be 
\begin{equation} %
\Lambda_{UR}(\beta) = \pi^{2/3} \ \beta \hbar c = \pi^{2/3} \ 
\frac{\lambda_C}{\tau} \; .   \label{lambdaUR}  \end{equation} % %

The pressure $p$ and the mass density $\rho$ in (\ref{Friedmann1}) and
(\ref{Friedmann2}) are those of matter and radiation present in the
Universe, introduced through their expressions for ideal
gases.  Interactions are only taken into account through reactions
supposed to take place in restricted conditions.  As examples, a weak
Thomson scattering lies behind thermal equilibrium before
recombination, and pair production will be responsible for the
existence of a huge number of electrons when $kT$ is higher than
$\approx 0.5 MeV$.

The values $\Lambda =0$, $k=0$ lead to very simple solutions and are
helpful in providing a qualitative idea of the general picture.  They
will be used as reference cases.  We shall later exhibit the expression of
 $H(z)$ in the pre--recombination period,
as well as the implicit expression of $a(t)$.  Nevertheless, in order to
get a firmer grip on the relevant contributions and the role of each
term, it is useful to review the customary discussion on the matter-
and radiation- dominated ages.

%%%%%%%%%%%%%%%%%
\subsection{Matter--dominated age}
%%%%%%%%%%%%%%%%%

Always in the standard approach, baryons (essentially nucleons)
dominate the energy content at present time.  This domination goes
back to the ``turning point'' time given below (equation
(\ref{turningpoint})), when radiation takes over.  Protons are
non-relativistic during all this period.  The standard argument runs
as follows.  Matter pressure has the expression $p_{b} = n_{b}
kT_{b}$.  It appears, however, always in the combination $\rho _{b} +
p_{b}/c^{2}$ = $n_{b}[m+kT_{b}/c^{2}]$ = $\frac{n_{b}}{c^{2}}
[mc^{2}+kT_{b}]$.  Thus, $p_{b}$ is negligible in the non-relativistic
regime.  Putting $p=0$, the reference case with $\Lambda =0$ and $k=0$
has for equations
\[
{\dot{H}}= - \;\frac{3}{2}\;H^{2} = - \;\frac{3}{2} 
\frac{\rho_{b}}{\rho_{crit}} \; .
\]
The general solution is 

\begin{equation}
\frac{1}{H\left( t\right) }=\frac{3}{2}\ t+C.  \label{dust2}
\end{equation}
Let us quickly examine $3$ models, two with $k = 0$, $\Lambda = 0$ 
and a third case with $\Lambda \ne 0$.
\vskip .5cm

%%%%%%%%%%%%%%%%%
{\bf{a) Matter--domination: dust Universe }} 
%%%%%%%%%%%%%%%%%

This unrealistic model  supposes matter domination all the time along. It 
takes at the ``beginning'' $H_{t=0}=\infty$. The integration constant 
$C$ vanishes and the solution is simply
$$ %
H(t)=\frac{2}{3t} \; . %
$$ %
This means that %
$$ %
\frac{d a}{a} = \frac{2}{3} \ \frac{dt}{t}\ \ . 
$$ %
The expressions relating the Hubble function, the expansion parameter,
the red-shift, and the density follow immediately (we reinsert $H_{0}$
for convenience):%
\begin{equation} \frac{H^{2}}{H_{0}^{2}} = (1+z)^{3}. 
	\label{Hmatterdom}
\end{equation} %
\[
\frac{a(t)}{a(t_{0})}=\left( \frac{t}{t_{0}}\right)^{2/3} \; ;
\]
\begin{equation} %
1+z=\left( \frac{t_{0}}{t}\right) ^{2/3} = \left(
\frac{2}{3H_{0}\;t}\right)^{2/3} \; ; \label{matterdomz}
\end{equation} %
\begin{equation} %
\frac{\rho_{b}}{\rho_{crit}}=\frac{H^{2}}{H_{0}^{2}} \ \Omega_{b0}  =
(1+z)^{3} \ \Omega_{b0} \; .\label{rhobb}
\end{equation} %
The age of the Universe can be got from (\ref{matterdomz}), by putting
$z = 0$.  One obtains $t_{0}$ = $2/(3 H_{0})$ $\approx 6.5 \times
10^{9}$ years, a rather small number.  There is, as said, a serious
flaw in this exercise-model: it supposes that matter dominates down to
$t \approx 0$, which is is far from being the case.  Furthermore, the
protons cannot, of course, be non-relativistic at the high
temperatures of the ''beginning'' and matter pressure should be added. 
Let us see two more realistic cases.
\vskip .5cm

%%%%%%%%%%%%%%%%%
%%%%%%%%%%%%%%%%%
{\bf{b) Matter--domination: present time}} 
%%%%%%%%%%%%%%%%%

Let us go back to the general solution (\ref{dust2}) and fix the
integration constant by the present value
$$%
\frac{1}{H\left( t_0\right) } = \frac{1}{H_0} =\frac{3}{2}\ t_0 +C \ .
$$%
The solution, now more realistic, will be
\begin{equation}
H(t) = \frac{H_0}{1 + \frac{3}{2} H_0 (t - t_0)} .\label{MD1}
\end{equation}
Integrations of $\frac{da}{a} = H(t) dt $ leads then to
\begin{equation}
1 + z = \frac{a_0}{a(t)} = \frac{1}{[1 + \frac{3}{2} H_0 (t - t_0)]^{2/3}} .
\label{MD2}
\end{equation}
Equation (\ref{rhobb}) keeps holding. 
Using that equality, and the value (\ref{rhocrit}) of the critical density, we find
\begin{equation} %
\rho _{b}=1.878\times 10^{-26}\; (1+z)^{3}\;\Omega _{b0}\;h^{2}\;
[kg\;m^{-3}] 
\end{equation} %
Dividing by  the proton mass, the number density is
\begin{equation} %
n _{b}=11.2 \times \;(1+z)^{3} \;\Omega _{b0}\;h^{2} \; \; [m^{-3}] 
\label{np}
\end{equation} %
Actually, $\Omega _{b0}$ = $1$ in the reference case we are
considering.  This gives a few nucleons per cubic meter at present
time.  The age of the Universe is basically the same as that for the
dust Universe: we look for the time $t$ corresponding to $z
\rightarrow \infty$, and find $t_0 - t$ = $2/(3 H_0)$.  Equations
(\ref{MD1}) and (\ref{MD2}) hold from the turning point down to
present times (provided $k = 0$ and $\Lambda$ = 0).
\vskip .5cm

%%%%%%%%%%%%%%%%%%%%%%%%%%%%
{\bf{c) Matter--domination: k = 0 but $\Lambda \ne$ 0 }} 
%%%%%%%%%%%%%%%%%%%%%%%%%%%

Recent evidence for $k = 0$ and a nonvanishing cosmological constant 
at present time gives to this case the prominent role.

Let us insert (\ref{zandH}) into  (\ref{rhobb}), to get 
$$%
\rho = \rho_0\  e^{-\ 3 \int_{t_0}^t H(t) dt}
$$
and then insert this expression into the Friedmann equation (\ref{Htwo}):
$$%
H^{2} = 2 \left( \frac{4 \pi G}{3}\right) \rho_0\ e^{-\ 3 \int_{t_0}^t
H(t) dt} + \frac{\Lambda c^{2}}{3} \; .
$$%
The time derivative gives
$$%
\frac{d H}{d t} = \frac{3}{2} \left(\frac{\Lambda c^{2}}{3} -
H^{2}\right) .
$$%
Integration leads then to the expression
\begin{equation}
H(t) = \sqrt{\frac{\Lambda c^{2}}{3}}\ \frac{\left(\sqrt{\frac{\Lambda
c^{2}}{3}} + H_0 \right) e^{3 \sqrt{\frac{\Lambda c^{2}}{3}} (t -
t_0)} - \left( \sqrt{\frac{\Lambda c^{2}}{3}} - H_0 \right)}{
\left(\sqrt{\frac{\Lambda c^{2}}{3}} + H_0 \right) e^{3
\sqrt{\frac{\Lambda c^{2}}{3}} (t - t_0)} + \left( \sqrt{\frac{\Lambda
c^{2}}{3}} - H_0 \right)} \ .\label{MD3}
\end{equation}
This expression for $H(t)$ gives $H = H_0$ when $t \rightarrow t_0$
and tends to (\ref{MD1}) when $\Lambda \rightarrow 0$.  To have it in
terms of more accessible parameters, we may rewrite it as
\begin{equation}
H(t) = H_0 \sqrt{\Omega_{\Lambda}}\
\frac{\left(\sqrt{\Omega_{\Lambda}} + 1 \right) e^{3 H_0
\sqrt{\Omega_{\Lambda}} (t - t_0)} - \left( \sqrt{\Omega_{\Lambda}} -1
\right)}{ \left(\sqrt{\Omega_{\Lambda}} + 1 \right) e^{3 H_0
\sqrt{\Omega_{\Lambda}} (t - t_0)} + \left( \sqrt{\Omega_{\Lambda}} -
1 \right)} \ \ .\label{MD32}
\end{equation}
To get the relation with $z$, we notice that 
$$%
H^{2} = 2 \left( \frac{4 \pi G}{3}\right) \rho_0 (1 + z)^3 +
\frac{\Lambda c^{2}}{3} = H_0^2 \left[\Omega_b (1 + z)^3 +
\Omega_{\Lambda}\right] 
$$%
gives
$$%
H^{2} - \frac{\Lambda c^{2}}{3} = 2 \left( \frac{4 \pi G}{3}\right)
\rho_0 (1 + z)^3
$$%
$$%
  H_0^{2} - \frac{\Lambda c^{2}}{3} = 2 \left( \frac{4 \pi
G}{3}\right) \rho_0  \ ,
$$%
which together imply
$$%
\frac{H^{2} - \frac{\Lambda c^{2}}{3}}{ H_0^{2} - \frac{\Lambda
c^{2}}{3}} = (1 + z)^3 \ .
$$%
Alternatively,
\begin{eqnarray}
(1 + z)^3 &=& \frac{H^2 - H_0^2\ \Omega_{\Lambda}}{H_0^2\ (1 -
\Omega_{\Lambda})} \ ; \\
\Omega_b + \Omega_{\Lambda} &=& 1 \ .
\end{eqnarray}
It remains to use (\ref{MD3}) to obtain
\begin{equation}
1 + z =  \left(4\ \frac{\Lambda c^{2}}{3}\right)^{1/3} 
\frac{e^{ \sqrt{\frac{\Lambda c^{2}}{3}} (t -
t_0)}}{\left[\left(\sqrt{\frac{\Lambda c^{2}}{3}} + H_0\right)\ e^{3
\sqrt{\frac{\Lambda c^{2}}{3}} (t - t_0)} +
\left(\sqrt{\frac{\Lambda
c^{2}}{3}} - H_0\right)\right]^{2/3}} \ ,
\end{equation}
which is the same as
\begin{equation}
\frac{a_0}{a(t)} = 1 + z =  \left(4\ \Omega_{\Lambda}\right)^{1/3} 
\frac{e^{H_0 \sqrt{\Omega_{\Lambda}} (t -
t_0)}}{\left[\left(\sqrt{\Omega_{\Lambda}} + 1\right)\ e^{3 H_0 
\sqrt{\Omega_{\Lambda}} (t - t_0)} +
\left(\sqrt{\Omega_{\Lambda}} - 1\right)\right]^{2/3}} \ .
\end{equation}

The neglected matter pressure is given by %
\begin{equation} %
\frac{p_{b}}{c^{2}\rho _{crit}}= 9.2 \times 10^{-14} \ T_{b} \ \left( 
1+z\right) ^{3} \ \Omega_{b 0} \;  . \label{neglected}
\end{equation} %
This will be one of our main points.  Matter pressure is neglected in
usual treatments for the reasons given above, which assume an ideal
gas.  We intend to take into account interactions between nucleons,
and shall find indications of a very abrupt raise of this pressure
during the PNS period.

%%%%%%%%%%%%%%%%%%%%%%%%%%	    
 \subsection{Radiation--dominated age}   %		 
%%%%%%%%%%%%%%%%%%%%%%%%%%
For photons, the temperature behaves as a frequency so that, by the 
very definition of red--shift, $T_{\gamma } = T_{\gamma 0} \ (1+z)$.  For 
example, hydrogen recombination takes place at $T_{\gamma } = T_{b} 
\approx 3000 K$.  This, together with the present value $T_{\gamma 0} 
\approx 2.7$ for the thermal background, gives a red--shift $(1+z) \approx  
1.1 \times 10^{3}$.  The mass--equivalent density is therefore %
\begin{equation} %
\frac{\rho _{\gamma}}{\rho _{crit}}=2.1 \times
10^{-7}\;T_{\gamma}^{4}\;h^{-2}=1.1 \times 10^{-5} \; 
(1+z)^{4}\;h^{-2} = \Omega_{\gamma 0} \; 
(1+z)^{4}\ ,  \label{rogamrocrit}
\end{equation} % 
where  
 \begin{equation} %
\Omega_{\gamma 0} = 1.1 \times 10^{-5} \;h^{-2} \; . \label{defN}
\end{equation} % %

Consider again the reference case $k = 0$, $\Lambda = 0$.  Because
$\epsilon_{\gamma} = 3\;p_{\gamma }$ and $\rho _{\gamma } =
\frac{\epsilon_{\gamma}}{c^{2}}$, we have
\[
{\dot H} = - \ 2 \ H^{2} = -\ 2 \  \frac{\rho _{\gamma }}{\rho _{crit}} 
\; , 
\]
leading automatically to  
\begin{equation} %
H^{2} = \Omega_{\gamma 0}  (1 + z)^{4} \; . \label{aggadois}
\end{equation} % %
Solving the equation is only necessary to fix the relation between the 
time parameter and the red--shift.  The solution, %
\[
H(t) = \frac{1}{2t} \; , \] %
implies %
 \begin{equation} %
 t = \frac{1}{2 \sqrt{\Omega_{\gamma 0}} (1 + z)^{2}} \ \; ; \; \; \frac{a(t)}{a_{0}} 
 = \Omega_{\gamma 0}^{1/4} \sqrt{2 t} \; .  \label{raddomz} \end{equation} % %

Before recombination (that is, for higher z's) there is thermal 
equilibrium between matter and radiation, because electrons are free 
and the mean free path of the photons is very small.  An estimate of 
the energy per photon at a certain $z$ can be obtained from 
$kT_{\gamma 0}= 2.3 \times 10^{-10} MeV $, which leads to
\begin{equation}
kT_{\gamma } = k T _{\gamma 0}  (1+z)=2.32 \times 10^{-10} (1+z) \; 
MeV.
\label{kTz}
\end{equation}
For example, an energy of $4 \ MeV$ corresponds to $z \approx 2 \times 
10^{10}$.  Thus, the thermalized state before recombination makes of 
$T_{\gamma }$, or its corresponding red--shift, the best time 
parameter.  We shall retain for later use the expressions  %
\begin{equation} \frac{p_{\gamma }}{c^{2}\rho _{crit}}= \frac{1}{3} \ 
\frac{\rho _{\gamma }}{\rho _{crit}} = 7.  \times 10^{-8}\;T_{\gamma
}^{4}\;h^{-2} = \frac{\Omega_{\gamma 0}}{3} \; (1+z)^{4}
\label{radpressure}
\end{equation} %
\begin{equation}
\frac{\rho _{\gamma }}{\rho _{b}}=2.31 \times 10^{-5}\;(1+z)\;
\Omega_{b 0}^{-1} \; . \label{radmatratio}
\end{equation} %
At recombination time, $\frac{\rho _{b}}{\rho _{\gamma }}\simeq 39.5
\;\Omega _{b0}$.  The scale parameter, and consequently the
red--shift, behave quite differently in a matter--dominated Universe
(\ref{matterdomz}) and in a radiation--dominated one (\ref{raddomz}). 
Equation (\ref{radmatratio}) shows that radiation becomes dominant at
high $z$'s.  The turning point, or change of regime, takes place when
$\rho
_{\gamma }\simeq \rho _{b}$, or %
\begin{equation}
1+z\simeq 4.3 \times 10^{4}\;\Omega _{b0}  \;  . \label{turningpoint}
\end{equation} %  %
This corresponds to $t \simeq 7.4 \times 10^{-8}\; \Omega_{b 0} 
^{-3/2}/H_{0}$ = $ 2.2 \times 10^{10}\;\Omega_{b 0}^{-3/2}$ sec.  When 
there is no thermalization, non-interacting matter pressure 
is negligible with 
respect to radiation pressure by a factor $\frac{p_{b}}{p_{\gamma 
}}\approx \frac{10^{-10}}{1+z} \ T$.  

With our proton-related variables, the temperature during the 
thermalized period preceding recombination will be
\begin{equation} %
\tau =  \frac{k T_{\gamma}}{m c^{2}} = 2.5 \times 10^{-13} (1 + z) 
\; . \label{tauinz}
\end{equation} % %
In terms of those variables we shall have, for example, %
\begin{equation} %
n_{\gamma }=3.8 \times 10^{-39} \ (1+z)^{3} \lambda_{p}^{-3} = 0.24 
\left(\frac{\tau}{\lambda_{p}}\right)^{3} 
\label{ngammainz}
\end{equation} % %
and
\begin{equation} %
n_{b} = 6.3 \times 10^{-9} \ \Omega_{b0} h^{2} \ 
\left(\frac{\tau}{\lambda_{p}}\right)^{3} \ . \label{nbintau}
\end{equation} % %

%%%%%%%%%%%%%%%%%%
   \subsection{The Friedmann solutions}
%%%%%%%%%%%%%%%%%%
 Let us now go back to the general equations (\ref{H2}) and 
 (\ref{Hpoint}).  Thermalization at $z 
> 10^{3}$ has important formal consequences.  
 Dependence of a single temperature makes it more convenient to use 
 the variable $z$, and much of the discussion can be made in terms of 
 energies, by using (\ref{kTz}). Using the units given in Eq.(\ref{cosmounits}), 
adding matter (\ref{rhobb}) and radiation (\ref{rogamrocrit}) densities 
and extracting from (\ref{omegas}) the value of $\Omega_{b 0}$,
Eq. (\ref{H2}) becomes
\begin{equation}
 H^{2} (z) = \Omega_{\gamma 0} (1+z)^{4} - \frac{k}{a_{0}^{2}} \ 
 (1+z)^{2}+\frac{\Lambda }{3} + (1+z)^{3}(1 + \frac{k}{a_{0}^{2}} 
 - \frac{\Lambda }{3}  - \Omega_{\gamma 0} ) \; .  \label{solsimple2}
\end{equation}
This   will be modified when interactions between nucleons are 
taken into account (see equation (\ref{solution2}) below).  In the 
reference case, %
\begin{equation}%
 H^{2}= \Omega_{\gamma 0} (1+z)^{4} + (1 - \Omega_{\gamma 0})
 (1+z)^{3} \; , \label{solsimpleref}
\end{equation} %
which reduces to (\ref{Hmatterdom}) when the last term dominates the 
right-hand side, and to (\ref{aggadois}) when the first 
term dominates.

All this can be see from another point of view. In fact, we 
 have $(1+z) H(t)= - \frac{dz}{dt}$.  We can consequently introduce 
 the function $f(z) = \frac{H^{2}(z)}{H^{2}_{0}}$, with $f(0) = 1$ and 
 $a(z) = \frac{a_{0}}{ 1+z}$.  In the units (\ref{cosmounits}), 
 equation (\ref{Hpoint}) becomes %
\begin{equation}%
(1+z)\;\frac{df}{dz}=3\;f+3\;\frac{p}{c^{2}\rho _{crit}}+ 
\frac{k}{a_{0}^{2}} \;(1+z)^{2}-\Lambda \ .\label{goodie} %
\end{equation} 
 Use now x = 1+z, with $f(x=1) = 1$.  As radiation pressure largely 
 dominates at all time, the only contribution to $p$ will come from 
 (\ref{radpressure}).  We shall take $3\;\frac{p}{c^{2}\rho_{crit}} = 
 \Omega_{\gamma 0} x^{4}$, with $\Omega_{\gamma 0}$ as given in 
(\ref{defN}), and rewrite the equation 
 as %
\begin{equation}%
x\;\frac{df}{dx} = 3\;f + \Omega_{\gamma 0} x^{4}+ \frac{k}{a_{0}^{2}} 
 \;x^{2}-\Lambda .  \label{avatar2} %
 \end{equation} %
 The solution is %
\begin{equation} %
 f\left( x \right) = \Omega_{\gamma 0} x^{4} - \frac{k}{a_{0}^{2}} \ 
 x^{2}+\frac{\Lambda }{3} + x^{3}(1+ \frac{k}{a_{0}^{2}} - 
 \frac{\Lambda }{3}  - \Omega_{\gamma 0} ) \; , \label{solsimple1}
\end{equation} % %
which is the same as (\ref{solsimple2}). This point of view will be of 
interest when the pressure of matter is added. This is due to the fact that 
pressure appears only 
in the equation (\ref{dotH}) for the derivative of $H$ and 
not in the expression (\ref{Htwo}) for $H^2$.

Things are not that simple for the expansion parameter, or for the
relation between $z$ and $t$.  Noting that $x = a_{0}/a$, we have $$a
\; \frac{da}{dt}=\sqrt{\Omega_{\gamma 0} a_{0}^{4} -k
a^{2}+\frac{\Lambda }{3} \
a^{4}+ a_{0}^{3} (1-\Omega_{\gamma 0}+k-\frac{\Lambda }{3} ) \ a},$$
whose solution is given by %
\begin{equation} %
t=t_{0} + \int_{a(t_{0})}^{a\left( t\right) } \frac{y \; dy }{\sqrt{ 
\Omega_{\gamma 0} a_{0}^{4} - ky^{2}+\Lambda y^{4}/3+y \left(1 - 
\Omega_{\gamma 0} + k a_{0}^{-2} - 
\Lambda/3 \right) a_{0}^{3} }} \; \; .  \label{integfora}%
 \end{equation} % %
Most aspects for $z > 10^{3}$ can be discussed in terms of the 
red--shift, and we shall have little use for the time variable.

%%%%%%%%%%%%%%%%%%%%%%%%%%%%%%%%%%
%					%
   \section{Microphysics at the PNS period}	%
%					%
%%%%%%%%%%%%%%%%%%%%%%%%%%%%%%%%%
\label{physics}
There are many interesting questions concerning the applicability of 
usual physical assumptions and consequent results in the period 
between $ z\approx 10^{15}$ ($kT\approx 400 \; GeV$) and $z\approx 
10^{10}$ ($kT\approx 4 \; MeV $).  They are worth a brief parenthesis, 
as they constitute the background to the discussion of the particular 
questions we shall be concerned with.

First, concerning Particle Physics, it may be that the usual treatment 
of cross--sections need revision, due to scarcity of space.  In that 
treatment, particles are supposed to start in an interaction--free asymptotic 
region, interact in some finite intermediate region, and finish the 
process as a free object in yet another asymptotic  domain.  The 
ingoing and outgoing flows are compared to define the cross-section.  
Total volume is supposed to be much larger than the volumes of the 
particles concerned.  Even if for decay rates and production processes 
with two initial particles (such as pair production) the volume factor 
cancels out \cite{Wei95}, such processes can be inhibited by the 
smallness of complete (configuration plus momentum) phase space and by 
final state interactions.  The usual elementary definition of the 
scattering matrix~\cite{BD64} involves a Dirac delta function in four 
momentum which actually assumes a very large volume.  This is a rather 
difficult question, but one which can in principle be theoretically 
solved.

 Other, deeper Particle Physics aspects are to be considered. 
 Nucleons are non-relativistic at $z \approx 10^{10}$, but highly
 relativistic at $z \approx 10^{15}$.  How far will a proton, when
 energy grows from $4 \; MeV$ to $400 \; GeV$, remain a proton ?  Is
 it true that at the energies involved in the period we can already
 talk of deconfined quarks and unshielded gluons~?  It is usually
 supposed that at high enough energies nucleons loose their identities
 and the system must be considered as a quark--gluon plasma
 \cite{TT99}.  The energy at which that happens, however, remains
 unknown.  The search for a signal of quark deconfinement, which has
 been actively looked for in the interval $15-200 \; GeV$ per proton,
 has not yet given a definitive answer~\cite{UH99}.  The experimental
 results are consistent with the presence of deconfinement, but do not
 exclude other interpretations.  This means that the signal is not
 unambiguous, is not conclusive~\cite{Sal99}.  What happens at still
 higher energies is simply not known, as the mechanisms of confinement
 and eventual deconfinement are as yet unclear even from the purely
 theoretical point of view~\cite{GGS99}.  We shall here suppose that
 protons keep their identities and examine the consequences.  We shall
 find, as we proceed backwards in time, that radiation protons produce
 a pressure blockage at energies of the order of hundreds of $MeV$,
 much too low for confinement to take place.

Concerning Thermodynamics and Statistical Mechanics aspects, the 
system should be treated, quite probably, as a finite system 
\cite{Hil94}.  Another point concerns fermions in general.  The Pauli 
principle can be seen as a consequence of an effective repulsive 
potential between kin fermions, whose range is the thermal wavelength.  
This wavelength decreases with temperature.  At the period under 
consideration, it will be very small for electrons, but large for 
non-relativistic protons.  What has been said of phase space for cross 
sections should be repeated in this context: pair production of 
protons, for example, would be inhibited by the presence of many 
protons in the final state.  

Anyhow, the first problem to be faced is much simpler.  The picture we 
have of the early Universe comes from inserting {\em ideal} fluid 
equations of state in the Friedmann equations.  In the PNS period, 
excepting the reactions leading to nucleosynthesis and their 
nuclei--breaking inverses, interactions are not taken into account at 
all, let alone the possibility of any abrupt behavior related to 
phase-transitions.  Thus, the first thing to be done would be to 
consider a real, interacting gas.  This is already a difficult enough 
task.  We shall make a first attempt by taking into account the size 
of the nucleons.  Protons and neutrons are, of course, the hardest 
known objects.  Instead of structureless, pointlike particles, we 
shall assume a hard--sphere gas.

A negative point is that the only known means to do it is by 
considering a hard-sphere potential, and potentials loose progressively 
their meaning when the particles represented become more and more 
relativistic.  Taking into account pair creation and annihilation 
become more and more necessary to avoid difficulties akin to the Klein 
paradox \cite{BD64}.  We shall do it as carefully as possible, in the 
hope of finding effects in the non-relativistic regime.  The nucleon 
size--effect will be simulated by a hard-sphere with the nucleon radius.  
What we should take for the proton radius is as yet a matter of 
controversy \cite{Kin90}, but the uncertainty between $0.80$ fermi and 
$0.86$ fermi is, of course, irrelevant for the gross estimate we have 
in view.  We shall use $0.8 \times 10^{-13}$  cm for the proton radius 
and undertake 
the usual backward path, from $z \approx 10^{10}$ to $z \approx 
10^{15}$.

%%%%%%%%%%%%%%%%%%%%%%%%%%%%%%%%%%
%					%
   \section{A solvable model for primeval close packing}	%
%					%
%%%%%%%%%%%%%%%%%%%%%%%%%%%%%%%%%%
\label{firstpacking}

 We have presented in the introduction a very rough estimate of the
 red--shift at which a blockage can take place, and proceed now to a
 somewhat more elaborate evaluation.  The result will be essentially
 the same as long as only the remnant protons are concerned.  This
 more refined approach provides, however, general formulae which can
 be applied also to the protons belonging to the radiation. 

The hard-sphere gas is one of the great unsolved problems of
Theoretical Physics.  The virial coefficients have been calculated
analytically only up to the fourth order (by Boltzmann).  Numerical
results exist for higher orders, but the virial series has, if any, a poor
convergence.  The best picture of the system is given by a computer
simulation refined through the use of Pad\'e approximants.  The
outcome is a curve for the equation of state \cite{ANSLM85}.  It shows
a clear phase transition (possibly two), in which the pressure grows
steeply to $\infty$.  The curve can be parametrized by an equation of
state of the type %
\begin{equation} %
p_{b} = \frac{n_{b} \; k T}{1 - n_{b}/n_{c}}, 
\label{eqstate1} \end{equation} % %
where $ n_{c} = \frac{\sqrt{2}}{D^{3}}$, $D$ being the sphere
diameter.  This equation can be qualitatively understood in the
``excluded volume'' approach to the hard sphere gas.  The canonical
partition function for an $N$-particle gas with interactions given by
a potential $V_{ij}$ = $V(|{\bf r}_{i}-{\bf
r}_{i}|)$ is %
$$%
Q_{N} = \frac{1}{\lambda^{3 N}} \int d^{3}r_{1} d^{3}r_{2} \ldots 
d^{3}r_{N} \ e^{- \beta \sum_{i=1}^{N-1} \sum_{j=i+1}^{N}V_{ij}} 
$$%
$$%
= \frac{1}{\lambda^{3 N}}  \int d^{3}r_{1} d^{3}r_{2} \ldots 
d^{3}r_{N} \prod_{i=1}^{N-1} \prod_{j=i+1}^{N}e^{- \beta V_{ij}} .
$$%
For a hard--sphere potential, the integrand vanishes every time it
happens that $|{\bf r}_{i}-{\bf r}_{j}| < D/2$ for any pair $(i, j)$
of particles.  Thus, it all amounts to forbid the regions with $|{\bf
r}_{i}-{\bf r}_{j}| < D/2$ for every pair, or to exclude the interior
of all the spheres.  As the potential simply vanishes outside the
spheres, the final picture is that of an ideal gas in a volume reduced
by the total volume of the $N$ spheres.  With the volume $v_{s}$ =
$\frac{\pi}{6} \ D^{3}$ for each sphere, the equation would be $p(V -
N v_{s})$ = $N kT$, or $p$ = $n kT/(1 - n v_{s})$.  There are actually
some geometric factors before $v_{s}$, as the excluded volume
increases at close packing.  In a configuration like \ 
\mbox{\raisebox{-2pt}{$\circ$\hspace{-2pt}{$\circ$}
\hspace{-14pt}{\raisebox{4pt}{$\circ$\hspace{-2pt}{$\circ$}}}}} \ 
each sphere actually excludes a cube of size $D$.  These geometric
factors have been found by the authors of \cite{ANSLM85}, the second
suggested phase transition corresponding to the still tighter packing
of type \mbox{\raisebox{-2pt}{$\circ$\hspace{-2pt}{$\circ$}
\hspace{-12pt}{\raisebox{3.7pt}{$\circ$}}}} .

Let us now apply (\ref{eqstate1}) to our problem.  The proton radius gives 
$D^{3} = 4.1 \times10^{-39} \; cm^{3}$, so that \ $
n_{c} = 0.345 \times 10^{39} \; cm^{-3} $.
Using (\ref{nb}), we find %
\begin{equation}%
\frac{n_{b}}{n_{c}}= 3.18 \times 10^{-44}\;(1+z)^{3}\;\Omega 
_{b0}\;h^{2}\; .  \label{ratiobc}
\end{equation} % %
The denominator in (\ref{eqstate1}) vanishes when 
$
1+z=\frac{3.1 \times 10^{14}}{\Omega _{b0}^{1/3}\;h^{2/3}}  $,
corresponding to
\[
kT \approx  \frac{73.6}{\Omega _{b0}^{1/3} \; h^{2/3}}\; GeV \; . 
\]
Thus, if $h = 0.7$ and $\Omega _{b0} = 0.03$, then $kT \approx 300\
GeV$.  The recently more favored value $\Omega _{b0} = 0.2$ would give
$kT \approx 160\ GeV$.  This is an indication of close-packing for the
protons existing today, but at the very high value $z$ $ \approx
10^{15}$.  The close-packing effect will actually take place at much
lower $z$ because of the pair--production effect, to be discussed in
the next section.  Notice anyhow that, even neglecting the
pair--produced protons, the proton mean free path would be $ \approx
(n_{c} \lambda_{P}^{2})^{-1}$, comparable to $\lambda_{P}$ itself.

Using ( \ref{neglected}), the baryon pressure (\ref{eqstate1}) will be 
given by %
\begin{equation}
\frac{p_{b}}{c^{2} \rho_{crit}} = 9.2 \times10^{-14} \left( 1 + z \right)
^{3} \; \Omega_{b0}\; T \ \frac{1}{1 - 3.18 \times10^{-44} \; (1 + z)^{3} \; \;
\Omega_{b0} \; h^{2}} \; .
\end{equation}
As there is thermal equilibrium at the period of interest, $T$  will be the
radiation temperature and consequently
\[
\frac{p_{b}}{c^{2}\rho _{crit}}=\allowbreak 2.48 \times 10^{-13} \; 
\Omega _{b0} \; \left( 1+z\right) ^{4} \; \frac{1}{1- 3.18 \times 
10^{-44}\;(1+z)^{3}\;\;\Omega _{b0}\;h^{2}} \; .
\]
The relation between this and the radiation pressure is 
\[
\frac{p_{b}}{p_{\gamma}}= 3.17 \times 10^{-8} \; \Omega _{b0}\;h^{2}\
\frac{1}{ 1- 3.18 \times 10^{-44}\;(1+z)^{3}\;\;\Omega _{b0}\;h^{2}}
\; .
\]
We can introduce the notation 
\begin{equation} %
M =  2.48 \times 10^{-13}\;\Omega _{b0} 
\;h^{2} , \; \ \ Q =  3.15 \times 10^{-15}\;\;\Omega 
_{b0}^{1/3}\; h^{2/3}  \label{MQ}
\end{equation} 
and write %
\begin{equation}  %
\frac{p_{b}}{c^{2}\rho _{crit}}=  M\left( 1+z\right) ^{4}\frac{1}{
1-[Q\;(1+z)]^{3}}\ . %
\end{equation} %  %
This is to be added to the Friedmann equation in the form
(\ref{avatar2}), which becomes
\begin{equation} %
x\;\frac{df}{dx}=3\;f+\Omega_{\gamma 0} x^{4}+ \frac{k}{a_{0}^{2}}
\;x^{2}+\allowbreak Mx^{4} \frac{1}{1-(Qx)^{3}}-\Lambda\ . 
\label{xfprime1} \end{equation} % %
The solution is %
$$ %
 f(x)=\Omega_{\gamma 0}
 x^{4}-\frac{k}{a_{0}^{2}}x^{2}+\frac{\Lambda}{3} +x^{3}\left[
 1+\frac{k}{a_{0}^{2}}-\frac{\Lambda}{3} -\Omega_{\gamma 0} \right]
 +\frac{M}{3} \; \times \; \; \; \; \; \; \; \; \; \; \; \; \; \; \;
 \; \; \; \; \; \; \; \; \; \; \; \;
$$ %
\begin{equation}   %
x^{3}\left\{ \ln \frac{(1-Q) \sqrt{ 1+Qx+Q^{2}x^{2}}}{(1-Qx)
\sqrt{1+Q+Q^{2}}}+\arctan \frac{\sqrt{ 3}\ Qx}{2+Qx}-\arctan
\frac{\sqrt{3}\ Q}{2+Q} \right\} .        \label{solution2}
\end{equation} % %
This is just (\ref{solsimple1}), with the additional term proportional
to $M$.  As $Q\ll 1$, a good approximation is %
$$ %
f(x)=\Omega_{\gamma 0}
x^{4}-\frac{k}{a_{0}^{2}}\ x^{2}+\frac{\Lambda}{3} +x^{3}\left[
1+\frac{k}{a_{0}^{2}}-\frac{\Lambda}{3} \right] \; \; \; \; \; \; \;
\; \; \; \; \; \; \; \; \; \; \; \; \; \; \; \; \; \; \; \; \; \; \;
\; \; \; \; \; \;\; \; \; \; \; \; \; \; \; \; \; \; \; \;
$$ %
\begin{equation}   %
\; \; \; \; \; \; \; \; \; \; \; \; \;  \; \; \; \; \; \; \; \; \; \; \; \; \; \;+\frac{M}{3} \;
x^{3}\left\{ \ln \frac{\sqrt{ 1+Qx+Q^{2}x^{2}}}{1-Qx} +\arctan
\frac{\sqrt{ 3}\ Qx}{2+Qx} \right\} .
\end{equation} % %
When $x$ tends to $Q^{-1}$ from smaller values, the functions $f(x)$
and $H(z)$ become infinite.  The implicit solution for $a(t)$ is still
given by (\ref{integfora}), with $f(x)$ now given by
(\ref{solution2}).

We arrive thus at the following provisional picture.  If we proceed
backwardly in time, there will be a value of the increasing red-shift
for which matter pressure produced by the remnant protons becomes
practically infinite.  This value corresponds to energies of a few
hundreds of GeV. Of course, at such energies the very notion of
potential is unacceptable.  But this is only a first step in our
reasoning chain.  Actually, there will be a much larger number of
protons in the medium.  We have up to now neglected those which are
produced in pairs by the radiation background.  If that number is high
enough, the pressure may become exceedingly high at lower energies, in
which potentials still do make sense.

%%%%%%%%%%%%%%%%%%%%%%%
   \section{Radiation-belonging nucleons}   %
%                                            %
%%%%%%%%%%%%%%%%%%%%%%%
\label{radionucleons}

The radiation background contains a large amount of massive particles
as soon as the pair production $\gamma \gamma \rightarrow e^{-} \
e^{+} $ threshold is attained.  As to the process of hadron production  \
$\gamma \gamma \rightarrow p {\bar p}$, it is clearly at work at $kT$
around $1\ GeV$.  Actually, the reaction threshold is much
lower~\cite{LL74}, because of the huge number of photons.  Even at
lower energies, there are many photons with energy high enough to
produce pairs.  The number of radiation-created hadrons becomes of the
same order of the number of photons a little above the threshold
\cite{ZN81}, much larger in effect than the number of remnant protons
considered previously.  Let us say a few more words on these
statements.

The basic question would be: when we proceed backwards in time, from
which energy on can we consider that the reaction \ \ $\gamma \gamma
\rightarrow p {\bar p}$ \ \ is in equilibrium ?  This energy is
important because, above it, the remnant protons are negligible and
the previous argument should be applied, instead, to the
radiation-belonging protons.

With an annihilation cross section $\sigma_{a} \approx 2 \times
10^{-26} cm^{2}$, the annihilation mean free path of a proton is
$\lambda_{a} $ $\approx$ $\frac{1}{n_{\gamma} \sigma_{a}} $ $\approx $
$10^{23} (1+z)^{-3}$ cm.  This gives $\approx $ $10^{-22} $ cm at $
400$ GeV, $\approx $ $2 \times 10^{-21} cm$ at $20 GeV$, and $10^{-7}$
cm at $4$ MeV. For the inverse reaction, pair creation, we could use
the Breit--Wheeler cross-section~\cite{BLP71}, but the exact value is
not necessary in our simplified approach.  An estimate, using only
mean free paths, can be made along the following lines.  The mean free
path for a photon, due to pair creation, is $\lambda_{\gamma \gamma
\rightarrow p {\bar p}}$ = $[n_{\gamma} \sigma_{\gamma \gamma
\rightarrow p {\bar p}}]^{-1}$.  This means that, on the average, a
$\gamma$ will meet another $\gamma$ to produce a pair every time the
volume $\lambda_{\gamma \gamma \rightarrow p {\bar p}} \
\sigma_{\gamma \gamma \rightarrow p {\bar p}}$ is spanned by a
$\gamma$.  Thus, a traveling photon will ``deposit'' one antiproton at
each volume of that value.  But that volume is just $1/n_{\gamma}$, so
that the concentration of antiprotons is roughly the same as that of
the photons.

A more precise description of the interplay between annihilation and
pair production requires a detailed analysis of the kinetics involved. 
A kinetic estimate~\cite{ZN81} gives $\tau \approx 1/44 \approx 0.02$
(corresponding to $\approx 20\ MeV$) for the temperature above which
there is equilibrium.  We insist that, as soon as chemical equilibrium
is attained, at kT $\approx 20\ MeV$, the number of antiprotons
becomes enormous, of the order of $n_{\gamma}$.

We can also estimate the temperature at which the number of
pair--produced protons (or antiprotons) becomes larger than the number
of remnant protons, by equating (\ref{npbar2}) and (\ref{nbintau}). 
The result of a numerical analysis is that the number of antiprotons
overcomes that of the remnant protons at $\tau \approx 0.058$.  Thus,
we can choose for security a reasonable value $\approx 0.06$
(corresponding to $\approx 60\ MeV$) and say that, for $\tau$ above
it, the reaction is in equilibrium and there will be a large amount of
protons and antiprotons.  At $k T \approx 60\ MeV$, these protons and
antiprotons are nonrelativistic and for them the potential--based
arguments are valid.

We have repeatedly said that the concentration of protons is actually
of the same order of magnitude of that of photons, many orders of
magnitude above the number of the remnant protons we have considered
in the previous section.  In fact, using the numerical factor
$n_{\gamma}/n_{\bar p} = 4/3$ discussed at the end of Appendix A, and
equations (\ref{ngammainz}) and (\ref{nbintau}), we find
\begin{equation} %
\frac{n_{\bar p}}{n_{b}} = 2.9 \times 10^{7} \ [\Omega_{b} 
h^{2}]^{-1} \ . \label{nbarpovernb}
\end{equation} % %  

There is consequently an abrupt jump in the concentration, and
(\ref{ratiobc}) will change dramatically.  To take this effect into
account, it is sufficient to change the numerical parameters
(\ref{MQ}): $M$ and $ Q^3$ must be multiplied by $\frac{n_{\bar p}}{
n_b}$, given by (\ref{nbarpovernb}).  Thus, the formulae of the
previous section can be used, with the parameters $M$ and $Q$ of
(\ref{MQ}) replaced by
\begin{equation} %
	M^{\prime} = \frac{n_{\bar p}}{ n_b} \ M = 7.2 \times 10^{-6} \; ;
\; Q^{\prime} = \left( \frac{n_{\bar p}}{ n_b} \right)^{1/3}\ Q = 9.56
\times 10^{-13} \; . %
\end{equation} %
In that case, $\frac{p^{\prime}_{b}}{c^{2}\rho _{crit}}= 
M^{\prime}\left( 1+z\right) ^{4}\frac{1}{1-[Q^{\prime}\;(1+z)]^{3}}$, 
or, as $M^{\prime}$ = $0.65 h^{2} \Omega_{\gamma 0}$, 
\begin{equation}  %
	\frac{p^{\prime}_{b}}{c^{2}\rho _{crit}} =  
\frac{1.95 h^{2}}{%
1-[Q^{\prime}\;(1+z)]^{3}} \frac{p_{\gamma}}{c^{2}\rho _{crit}} =
\frac{0.65 h^{2}}{
1-[Q^{\prime}\;(1+z)]^{3}} \ \frac{\epsilon_{\gamma}}{c^{2}\rho _{crit}} .  
\end{equation} %  %
The term $\Omega_{\gamma 0} x^{4}$ in (\ref{xfprime1}) is the
radiation contribution.  That equation becomes
\begin{equation} %
(1+z)\;\frac{df}{d(1+z)}=3\;f+ \frac{k}{a_{0}^{2}}
\;(1+z)^{2}-\Lambda + \Omega_{\gamma 0} (1+z)^{4} \left[1 + \frac{
0.65\ h^{2}}{ 1-[Q^{\prime}\;(1+z)]^{3}}\right] .
\label{xfprime2} %
\end{equation} % %
The last term corresponds to the radiation contribution, with the term
in $h^{2}$ giving the nucleon interaction correction to the equation
of state (\ref{radpressure}).  The solution is still
(\ref{solution2}), but with the replacements $M \rightarrow
M^{\prime}$ and $Q \rightarrow Q^{\prime}$.

Matter pressure now becomes infinite at $z = 1.0 \times 10^{12}$, or
$kT = 232 MeV$.  At such energies, it is possible that
deconfinement, conjectured to happen at $\approx 150 MeV$, has taken place.
 We are, however,
 neglecting some effects (discussed below) which would tend to lower that value. 
And even at $\approx 230 MeV$ the 
non--relativistic argument, based on the notion of a potential, is valid.
  Notice that the energy at which pair production attains 
equilibrium is much lower:  $\tau \approx 0.05$ corresponds  to
$k T_{\gamma} \approx 50 \ \ MeV$. A last point is that antiprotons 
attain a concentration comparable to that of photons a bit above the 
reaction threshold, which can be lower than the equilibrium 
temperature. We are, thus, quite probably overestimating the above 
energies.

%%%%%%%%%%%%%%%%%%%%%%%%%%%%
%											 
	\section{Conclusions and speculations}	 %
%											 
%%%%%%%%%%%%%%%%%%%%%%%%%%%%

The conclusion is that, as soon as pair creation starts up, the number
of protons becomes so high as to create a blockage.  This might mean
that the reaction is inhibited by a ``wall'' of occupied phase space
in  the final state, or that the use of cross--sections is inadequate, or
still that usual thermodynamics does not apply.  It is common
knowledge that, in consequence of fundamental requirements, pressure
in a fluid cannot go to infinity.  It can at most attain the
incompressibility limit, at which the equation of state is $p =
\epsilon$ and beyond which causality would be violated.  With $h =1$,
this point would correspond to $z \approx 10^{12}$.

In its details, the primeval blockage we have dealt on has,
unfortunately, a crucial dependence on the values of the as yet
undetermined cosmological parameters.  Some of the neglected aspects
would, however, add to the repulsion and, consequently, to the effect. 
Notice that any effect leading to a higher effective value for the
sphere volumes in (\ref{eqstate1}) will bring the critical redshift to
lower values.  For instance, scattering theory will tell us that a
sphere appears larger in the quantum case: na\"{\i}ve shadow
scattering by a disk of radius $r$ will see a transverse area $4 \pi
r^{2}$, instead of the classical $\pi r^{2}$.  This would mean an
effective radius twice as large, and a red-shift half that found above
($z = 0.5 \times 10^{12}$, or $kT = 116\ MeV$).  Still another
possiblility is the breaking of Debye electromagnetic screening, which
would lead to an increase of the electrostatic Coulomb repulsion. 
Finally, we have completely neglected the Fermi repulsion due to Pauli
exclusion.  This effect is feeble in the ultrarelativistic regime, but
the thermal wavelengths (\ref{thermlength}) or (\ref {lambdaNR}) can
be larger than the diameter of the excluded sphere for the range of
energies found.

We have not attempted to relate our results to the current
negative--pressure approach to inflation \cite{McC51,Mur73}.  We shall
only indulge in a wild remark.  As Lee and Yang have taught us long
ago~\cite{LY52}, equations of state have the same analytical form both
sides of a phase transition.  If equation (\ref{eqstate1}) holds for
$n_{b}/n_{c} > 1$, the pressure will be negative ``the other side'' of
the thermodynamic singularity.  In effect, the expression for the
total radiation pressure, including the
nucleon--antinucleon pairs, %
\begin{equation} %
\frac{p_{\gamma}}{c^{2}\rho _{crit}} = \Omega_{\gamma 0} (1+z)^{4}
\left[\frac{1}{3} + \frac{ 0.65 \ h^{2}}{
1-[Q^{\prime}\;(1+z)]^{3}}\right] \ ,
\end{equation} %
or (using $\epsilon_{\gamma}$ for the {\em ideal} energy density)
\begin{equation} %
p_{\gamma} = \left[\frac{1}{3} + \frac{ 0.65 \ h^{2}}{
1-[Q^{\prime}\;(1+z)]^{3}}\right] \; \epsilon_{\gamma} =
\left[\frac{1}{3} + \frac{ 0.65 \ h^{2}}{ 1-\left[\frac{k T_{\gamma}
(MeV)}{232}\right]^{3}}\right] \; \epsilon_{\gamma} \ ,
\label{eqstate}
\end{equation} %
will be negative in a short interval before the singular red--shift. 
A comparison of (\ref{H2}) and (\ref{solsimple2}) shows that, for the
large values of $z$ we are considering now, the energy density is
dominated by the quartic term.  Consequently, the ideal
$\epsilon_{\gamma}$ of the last expressions is the true radiation
energy and (\ref{eqstate}) is indeed an equation of state.  The equation
would be of the exponential inflationary type $p = - \ \epsilon$ \
type~\cite{WOL92} only for a very particular value of $z$.  We can
however, speculate on the possibility of a more general type of
inflation.  For a barotropic equation $p = (\gamma - 1) \epsilon$ an
extended, power--law type inflation~\cite{LM85} occurs for $\gamma$ in
the range $0 \le \gamma \le 2/3$~\cite{Ell90}.  For $h = 1$, this
would correspond to a tiny interval $1.14 \le Q^{\prime}\;(1+z) \le
1.25$ before the pressure singularity.

\section*{Acknowledgements}

We are deeply grateful to Professor R.A. Salmeron, of the \'Ecole
Polyt\'echni\-que, Palaiseau, and CERN, Geneva, for illuminating
discussions on the experimental evidence concerning quark
deconfinement.  One of the authors (R.A.)
is grateful to FAPESP (S\~ao Paulo, Brazil) for financial support. 
\appendix

\section{Relativistic Gases}
\label{Relativistic Gases}

We justify here some statements and formulae of the text, and rewrite some
of the most usual expressions in units specially adequate to our case.

To see how (\ref{thermlength}) comes out, let us recall that the 
grand-canonical partition function for a gas of non-interacting 
quantum particles with chemical potential $\mu$ is  the trace of the 
density operator: 
$$
\Xi(V,\beta,\mu) = tr \left[ e^{- \beta \sum_{i} (\epsilon_{i} - \mu) {\hat 
n}_{i}} \right] = \sum_{\{n_{j}\}} \langle n_{0} n_{1} n_{2} \ldots | 
e^{- \beta \sum_{i} (\epsilon_{i} - \mu) {\hat 
n}_{i}} |  n_{0} n_{1} n_{2} \ldots \rangle =
$$
$$
\sum_{n_{0}}  \sum_{n_{1}}  \sum_{n_{2}}  \ldots e^{- \beta 
(\epsilon_{0} - \mu) n_{0}}  e^{- \beta 
(\epsilon_{1} - \mu) n_{1}}  e^{- \beta 
(\epsilon_{2} - \mu) n_{2}} \ldots  =
 \prod_{i}  \sum_{n} e^{- \beta (\epsilon_{i} - \mu) n}  
 $$%
 $$%
 = \prod_{\epsilon} \; \sum_{n} e^{- \beta (\epsilon - \mu) n} .
$$
In this non-interacting case, each level contributes an 
independent factor.  The system can have also internal degrees of 
freedom, which will likewise contribute separately.  Suppose a single 
degree of freedom (spin, for example) taking $g$ possible values.  
The partition function will be %
\begin{equation} %
\Xi(V,\beta,\mu) = \prod_{\epsilon} \; \left[ \sum_{n} e^{- \beta 
(\epsilon - \mu) n} \right]^{g} \; . \label{grandpart3}  \end{equation} % %

The kind of statistics appears in the summation, which is over the
possible values of the occupation number $n$, from $n = 0$ up to the
maximum number of particles allowed in each state: $1$ for fermions,
$\infty$ for bosons.  To treat bosons and fermions at the same time,
we adopt the usual convention: upper signs for bosons, lower signs for
fermions.  Transforming the product into a summation by using the
formal identity $\prod_{\epsilon} \{ \ldots \} = \prod_{\epsilon}
[\exp(\ln \{ \ldots \} )] = exp \sum_{\epsilon} \ln \{ \ldots \}$, the
above expressions lead to %
\begin{equation} %
\ln \Xi^{B,F}(V,\beta,\mu) = \mp  g \sum_{\epsilon} \ln \left[ 1 \mp e^{- 
\beta(\epsilon - \mu) } \right]  \; .
\end{equation} % %
It is convenient to use the fugacity variable, either the usual 
non-relativistic fugacity $ z = e^{\beta \mu}$ or its relativistic 
version $ Z = e^{\beta \mu_R} = z e^{\beta m c^2} $.  If we do not care 
about zero--energy states, the sum over the energy 
levels can be replaced by an integral over the momenta through the 
prescription $ \sum_{\epsilon} \rightarrow h^{-3} \int d^{3}x 
d^{3}p$, which leads to 
$$
\ln \Xi^{B,F}(V,\beta,\mu) = \mp g \frac{4 \pi V}{h^{3}} \int_{0}^{\infty} 
p^{2} dp \ln \left[ 1 \mp Z e^{- \beta (p^{2} c^{2} + 
m^{2} c^{4})^{1/2}} \right]  \; .
$$
Expanding the logarithm and collecting like terms, the 
partition function acquires the form  %
\begin{equation} 
\hspace{-15pt}\Xi^{B,F}(V,\beta,z) = \exp \left\{ \frac{g V}{h^{3}}
\sum_{j=1}^{\infty} \frac{(\pm 1)^{j-1}}{j} z^{j} \int d^{3} p e^{- j
\beta [(p^{2} c^{2}
+ m^{2} c^{4})^{1/2} - m c^{2}]} \right\} .  \label{grandpart6} %
\end{equation} %  %
The relativistic thermal wavelength (\ref{thermlength}) appears now 
in the form %
\begin{equation} %
\frac{1}{h^3} \; \int d^3p\; \ e^{- \beta [(p^2 c^2 + m^2 c^4)^{1/2} - 
m c^2]} = \frac{1}{\Lambda_{T}^3(\beta)} , \label{integral} %
\end{equation} % %  
and the final expression for the grand-canonical partition 
function for a gas of non-interacting quantum particles is %
\begin{equation} %
\Xi ^{B,F} (V,\beta,z) = \exp \left\{ g V \; \sum_{j=1}^{\infty} 
\frac{(\pm 1)^{j-1}}{j} z^{j} \frac{1}{\Lambda_{T}^{3}(j \beta)} \right\} 
\; ,  \label{grandpart8} \end{equation} % %
or its  equivalent %
\begin{equation} %
\Xi ^{B,F} (V,\beta,z) = \exp \left\{ g 
\frac{4 \pi V}{h^{3} c^{3}} \; \frac{(m c^{2})^{2}}{\beta} \; 
\sum_{j=1}^{\infty} \frac{(\pm 1)^{j-1}}{j} Z^{j} K_{2} (j \beta m 
c^{2}) \right\} \; .
\label{grandpart9} 
\end{equation} % %
Here $K_{2}(x)$ is the modified Bessel function of second order.  
Limits can be found by using the properties %
$$ K_{2}(x) \approx 
\sqrt{\frac{\pi}{2 x}} e^{-x} \left(1 + \frac{15}{8 x} + \ldots 
\right) ; %
\; 
$$%
$$%
K_{1}(x) \approx \sqrt{\frac{\pi}{2 x}} e^{-x} \left(1 + 
\frac{3}{8 x} + \ldots \right) \ {\rm for \ x} >> 1;
$$ %
$$ K_{2}(x) \approx 
2 x^{-2} \; ; K_{1}(x) \approx x^{-1} \ {\rm  for \ x} << 1.
$$
$K_1(\beta m c^2)$ will appear only  in the energy expression. 
The non-relativistic and the ultra--relativistic limits give 
(\ref{lambdaNR}) and (\ref{lambdaUR}).
The pressure and the particle number follow by standard 
thermodynamic relations: %
\begin{equation} %
p V = k T \ln \Xi = g  k T  \sum^\infty_{l=1} \frac{(\pm)^{l-1}}{l} z^l 
\frac{1}{\Lambda_{T}^3(l \beta)},  \end{equation} 
$$ %
{\bar N} = \left[z \frac{\partial}{\partial z} \ ln \ \Xi (V,\beta,z) 
\right]_{V,\beta}  
= g \sum_{\epsilon} \; \frac{1}{z^{-1} e^{\beta \epsilon} \mp 
1}  \; \; \; \; \; \; \; \; \;\; \; \; \; \; \; \; \; \; \; \; \; \; \; \; \; \; \; \; 
\; \; \; \;\; \; 
$$ %
\begin{equation} %
\; \; \; \; \; \; \; \; \; \; \; \; \; \; \; 
=  \frac{g V}{h^{3}} \int \frac{d^{3}p}{z^{-1} e^{\beta 
\sqrt{p^{2} c^{2} + m^{2} c^{4}}} \pm 1} = g V \sum^\infty_{l=1}
(\pm)^{l-1} z^l  \frac{1}{\Lambda_{T}^3(l \beta)} \; . 
\end{equation} % %
The expressions in terms of integrals or of series are more or less
convenient, depending on the application in view.  We can extract the
density number of particles at energy $\epsilon$, $n _{\epsilon}$ =
$g \ [z^{-1} e^{\beta \epsilon} \mp 1]^{-1}$.  The average energy,
including the masses, is %
$$ %
 {\bar E} = - \left(\frac{\partial}{\partial
 \beta} \; \ln \Xi(V,\beta,z) \right)_{Z,V} = \sum _{\epsilon} n
 _{\epsilon} \epsilon 
\; \; \; \; \; \; \; \; \;\; \; \; \; \; \; \; \; \; \; \; \; 
\; \; \; \; \; \; \; \; \;\; \; \; \; \; \; \; \; \; \; \; \; 
\; \; \; \; 
$$ % 
 \begin{equation} %
 \; \; \; \; \; \; \; \; \; \; = 3 p V + \ 4 \pi \ g \left(\frac{m
 c^2}{\lambda_C^3}\right) \left(\frac{kT}{m c^2}\right)
 \sum_{l=1}^\infty \frac{(\pm)^{l-1}}{l} z^l e^{l \beta m c^2} K_1(l
 \beta m c^2) \ .  \end{equation} % %
The degree of degeneracy is 
\begin{equation} %
\hspace{-30pt} d = \frac{{\bar N} \Lambda_{T}^3(\beta)}{V} = g
\sum^\infty_{l=1} (\pm)^{l-1} z^l
\frac{\Lambda_{T}^3(\beta)}{\Lambda_{T}^3(l \beta)} = g
\sum^\infty_{l=1} \frac{(\pm)^{l-1}}{l} z^l \frac{e^{- \beta m c^2}
K_2(l \beta m c^2)}{e^{- l \beta m c^2} K_2(\beta m c^2)} \ .
\label{degeneracyindex}%
\end{equation} % %
We are particularly interested in $n = \frac{\bar N}{V}$.  For a
massless particle, the change of variables $x = p c/kT$ can be used
directly to give %
\begin{equation} %
 n = \frac{g}{2 \pi^{2}} \ 
\frac{\tau^{3}}{\lambda^{3}_{C}} \int_{0}^{\infty} \frac{x^{2} 
dx}{z^{-1} e^{x } \pm 1} \; .  \end{equation} % %

For a gas of photons (using g = 2, z = 1),
there are many expressions of interest: %
\begin{equation} %
n_{\gamma} = \frac{{\bar N}_{\gamma}}{V} = \frac{2}{h^{3}} \int 
\frac{d^{3}p}{ e^{\beta p c} - 1} = \frac{1}{\pi^{2}} \ 
\frac{\tau^{3} }{\lambda_{C}^{3}} \int_{0}^{\infty} \frac{x^{2} 
dx}{e^{x} - 1} \label{ngamma1} = 2 \sum^\infty_{l=1} 
\frac{1}{\Lambda_{UR}^3(l \beta)} 
\end{equation} % %
\begin{equation} %
= 2 \ 
\frac{\tau^{3}}{\pi^{2} \lambda_{C}^{3}} \sum^\infty_{l=1} 
\frac{1}{l^3} 
= \frac{2 \ \zeta(3)}{\pi^{2}} \left( \frac{k T_{\gamma}}{\hbar 
c}\right)^{3} = 0.244 \ 
\frac{\tau^{3}}{\lambda_{C}^{3}} \; .  \label{ngamma4} %
\end{equation} % %
The pressure is found to be $p_{\gamma} = \frac{\zeta(4)}{\zeta(3)} \
n_{\gamma} k T_{\gamma}$ = ${\bar E}_{\gamma}/3 V$ =
$\epsilon_{\gamma}/3$.  For a gas of fermions with g = 2 (like
protons or antiprotons), %
\begin{equation} %
 n_{\bar p} = \frac{{\bar N}_{p}}{V} = 2
\sum^\infty_{l=1} (-)^{l-1} z^l \frac{1}{\Lambda_{T}^3(l \beta)} =
n_{\bar p} = \frac{1}{ \pi^{2}} \
\left(\frac{\tau}{\lambda_{p}}\right)^{3} \int_{0}^{\infty}
\frac{x^{2} dx}{z^{-1} e^{\sqrt{1/\tau^{2} + x^{2} }} + 1}
\label{npbar2} \; .
\end{equation} % %
In the ultrarelativistic regime this becomes %
\begin{equation} %
n_{\bar p} = \frac{1}{\pi^{2}}  \ \frac{\tau^{3} }{\lambda_{C}^{3}} 
\int_{0}^{\infty} \frac{x^{2} dx}{e^{x} + 1} = \frac{3}{2} \  
\frac{\zeta(3)}{\pi^{2}} \ \frac{\tau^{3} }{\lambda_{C}^{3}} = 
0.183 \ \frac{\tau^{3} }{\lambda_{C}^{3}} \; .
\label{npbarUR}
\end{equation}  %
A factor $\frac{n_{\gamma}}{n_{\bar p}} = \frac{4}{3}$ comes from the
fermion repulsion effect, encapsulated in the sign in the integrand
denominator, opposite to that in (\ref{ngamma1}).  It is more
difficult to pack fermions than bosons together.  This can be seen
also from the limits of the degeneracy index (\ref{degeneracyindex}). 
In this ultrarelativistic regime, its values are, for photons and
antiprotons, respectively, $n_{\gamma} \ \Lambda^{3}_{UR}$ = $0.244 \
\pi$ = $0.766$ and $n_{\bar p UR} \ \Lambda^{3}_{UR}$ = $0.183 \ \pi$
= $0.575$.  The relativistic or nonrelativistic character of the
protons depend on the above numerical factors.  They are possibly
irrelevant to rough estimates of the main text, but may come to be
important in a more detailed consideration of Fermi repulsion.

\end{document}